\documentclass[final,5p,times,twocolumn,sort&compress]{elsarticle}
\usepackage{epsfig}

\usepackage{amssymb}
\usepackage{amsmath}

\journal{Physics Letters A}

\begin{document}

\begin{frontmatter}



\title{Magnetic Gr\"uneisen parameter and magnetocaloric properties \\
of a coupled spin-electron double-tetrahedral chain}
\tnotetext[t1]{This work was financially supported by the grant of the Slovak Research and Development Agency under the contract \mbox{No. APVV-0097-12.}}
 \author[label1]{Lucia G\'{a}lisov\'{a}}
 \ead{galisova.lucia@gmail.com}
 \author[label2]{Jozef Stre\v{c}ka}
 \ead{jozef.strecka@upjs.sk}
 \address[label1]{Department of Applied Mathematics and Informatics,
             Faculty of Mechanical Engineering, Technical University,
             Letn\'{a}~9, 042 00 Ko\v{s}ice, Slovakia}
 \address[label2]{Department of Theoretical Physics and Astrophysics,
             Faculty of Science, P.~J.~\v{S}af\'{a}rik University,
             Park Angelinum 9, 040 01 Ko\v{s}ice, Slovakia}

\begin{abstract}
Magnetocaloric effect in a double-tetrahedral chain, in which nodal lattice sites occupied by the localized Ising spins regularly alternate with three equivalent lattice sites available for mobile electrons, is exactly investigated by considering the one-third electron filling and the ferromagnetic Ising exchange interaction between the mobile electrons and their nearest Ising neighbours. The entropy and the magnetic Gr\"uneisen parameter, which closely relate to the magnetocaloric effect, are exactly calculated in order to investigate the relation between the ground-state degeneracy and the cooling efficiency of the hybrid spin-electron system during the adiabatic demagnetization.
\end{abstract}


\end{frontmatter}


\section{Introduction}
\label{sec:1}
The magnetocaloric effect (MCE), which is characterized by an adiabatic change of the temperature (or by an isothermal change of the entropy) under the variation of the applied  magnetic field, has a long history in cooling applications at various temperature regimes~\cite{War81}. Since the first successful experiment of the adiabatic demagnetization performed in 1933~\cite{Gia33}, the MCE is a standard technique for achieving the extremely low temperatures~\cite{Str07}.

The theoretical prediction and description of materials with an enhanced or even giant MCE create real opportunities for the effective selection of the construction for working magnetic refrigeration devices. Of particular interest is the investigation of the MCE in various one-dimensional (1D) quantum spin systems~\cite{Der06,Hon09,Can09,Tri10,Lan10,Top12,Kass13,Gal14,Str14,Zar} or hybrid spin-electron models~\cite{Per09,Gal15a,Gal15b}. The reason is a possibility of obtaining the exact analytical or numerical results as well as a potential use of these models for the explanation of  MCE data measured for real magnetic compounds. In particular, 1D models may give correct quantitative description of real three-dimensional (3D) magnets, when appropriate rescaling of material parameters are taken into account~\cite{Sol97,Has08,Kur10,Mat12}.

In the present Letter, we will investigate the MCE in a hybrid double-tetrahedral chain composed of the localized Ising spins and mobile electrons, which is exactly solvable by combining the generalized decoration-iteration mapping transformation~\cite{Fis59,Syo72,Roj09,Str10} and the transfer-matrix technique~\cite{Kra44,Bax82}. As has been shown in our previous works~\cite{Gal15a,Gal15b}, the considered 1D spin-electron system provides a suitable prototype model for theoretical investigation of the relation between the ground-state degeneracy and the cooling efficiency of the system during the adiabatic (de)magnetization.

The Letter is organized as follows. In Section~\ref{sec:2}, we will describe the investigated spin-electron double-tetrahedral chain and then, the basic steps of an exact analytical treatment of the model will briefly be recalled. Exact calculations of the basic thermodynamic quantities, such as the Gibbs free energy, the total magnetization, the entropy and the magnetic Gr\"uneisen parameter, will be realized in this section. In Section~\ref{sec:3}, we will particularly discuss the numerical results acquired for the ground state, the entropy and the magnetic Gr\"uneisen parameter as functions of the applied magnetic field under the assumption of the one-third electron filling of each triangular cluster and the ferromagnetic exchange interaction between the mobile electrons and their nearest Ising neighbours. Finally, the Letter ends up with a summary of our findings in Section~\ref{sec:4}.

\section{Model and its exact solution}
\label{sec:2}
Let us consider a magnetic system on a double-tetrahedral chain, where nodal lattice sites occupied by the localized Ising spins regularly alternate with three equivalent lattice sites available to two mobile electrons.
\begin{figure}[th!]
\begin{center}
\hspace{0.0cm}
\includegraphics[angle = 0, width = 0.75\columnwidth]{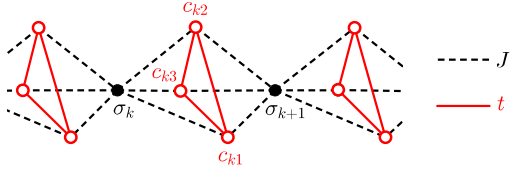}
\vspace{-0.1cm}
\caption{\small A part of the spin-electron system on the double-tetrahedral chain.  Full circles denote nodal lattice sites occupied by the localized Ising spins, while the empty
circles forming triangular clusters are available to mobile electrons.
}
\label{fig1}
\end{center}
\vspace{-0.5cm}
\end{figure}
The magnetic structure of the considered 1D spin-electron model is schematically illustrated in Fig.~\ref{fig1}. Assuming the exchange interactions between the nearest neighbours, the on-site Coulomb repulsion $U\geq0$ between two electrons of opposite spins occupying the same lattice site and the action of the external magnetic field $B$ on the mobile electrons and the localized Ising spins, the total Hamiltonian of the model reads:
\begin{eqnarray}
\label{eq:H_tot1}
{\cal H}\!\!\!&=&\!\!\!
-\,t\sum_{\langle i, j\rangle}\sum_{s\in\{\uparrow,\downarrow\}}
\left(c_{i,s}^{\dagger}c_{j,s}+\,c_{j,s}^{\dagger}c_{i,s}\right) \nonumber \\
        & &\!\!\!
				+\, \frac{J}{2}\sum_{\langle j, k\rangle}\left(n_{j,\uparrow}-n_{j,\downarrow}\right)\sigma_{k}^z  + U\sum_{j}n_{j,\uparrow}n_{j,\downarrow}\nonumber \\
        & &\!\!\!
-\, \frac{H_e}{2}\sum_{j}\left(n_{j,\uparrow}-n_{j,\downarrow}\right) -\, H_{I}\sum_{k}\sigma_{k}^z.
\end{eqnarray}
In above, the summation $\langle i, j\rangle$ runs over the lattice sites forming triangular clusters, while the summation $\langle j, k\rangle$ runs over the lattice sites of triangular clusters and the nearest-neighbouring nodal lattice sites. The operator $c_{i,s}^{\dagger}$ ($c_{i,s}$) represents usual fermionic creation (annihilation) operator for mobile electrons occupying the $i$th lattice site with the spin $s\in \{\uparrow,\downarrow\}$, $n_{j,s}=c_{j,s}^{\dagger}c_{j,s}$ is the number operator of the mobile electron at the $j$th lattice site and $\sigma_{k}^z$ labels the localized Ising spin at the $k$th nodal lattice site. The hopping parameter $t>0$ takes into account the kinetic energy of mobile electrons delocalized over triangular clusters and $J$ stands for the Ising-type coupling between mobile electrons and their nearest Ising neighbours. Finally, the last two terms in Eq.~(\ref{eq:H_tot1}) represent the Zeeman's energies of the mobile electrons and the localized Ising spins, which may be in general different due to a difference in the relevant Land\'e g-factors absorbed into the definition of the 'effective' magnetic fields $H_e = g_e \mu_{\rm B} B$ and $H_I = g_I \mu_{\rm B} B$ ($g_e$ is Land\'e g-factor of the mobile electrons, $g_I$ is Land\'e g-factor of the localized Ising spins and $\mu_{\rm B}$ is Bohr magneton).

\subsection{Partition function}
\label{subsec:21}

One can note that the spin-electron model under consideration can alternatively be viewed as the spin-$1/2$ Ising linear chain, whose bonds are decorated by triangular clusters available for two mobile electrons. From this point of view, the partition function of the system can exactly be derived within the generalized decoration-iteration mapping transformation~\cite{Fis59,Syo72,Roj09,Str10} (as described in our previous work~\cite{Gal15b}). As a result, one obtains a simple relation between the partition function ${\cal Z}$ of the investigated spin-electron tetrahedral chain and the partition function ${\cal Z}_{I}$ of the uniform spin-$1/2$ Ising linear chain with the effective nearest-neighbour coupling $J_{e\!f\!f}$ and the effective magnetic field $H_{e\!f\!f}$ :
\begin{eqnarray}
\label{eq:DIT}
{\cal Z}\left(\beta, J, t, U, H_{I}, H_{e}\right)= A^N{\cal Z}_{I}(\beta, J_{e\!f\!f}, H_{e\!f\!f}),
\end{eqnarray}
where $\beta=1/T$ is the inverse temperature (we set the Boltzmann's constant $k_{\rm B}=1$) and  $N$ is the total number of the nodal lattice sites (the localized Ising spins). The explicit expessions of the mapping parameters $A$, $J_{e\!f\!f}$ and $H_{e\!f\!f}$ emerging in Eq.~(\ref{eq:DIT}) can be obtained from the 'self-consistency' condition of the applied decoration-iteration transformation (see Eqs.~(9) and (10) in Ref.~\cite{Gal15b}). At this stage, the exact calculation of the partition function ${\cal Z}$ of the spin-electron tetrahedral chain is formally completed, because the partition function ${\cal Z}_{I}$ of the spin-$1/2$ Ising linear chain in a magnetic field is known~\cite{Kra44,Bax82}.

\subsection{Magnetization, entropy and magnetic Gr\"uneisen parameter}
\label{subsec:22}

In this part, we present the exact solution for the Gibbs free energy ${\cal G}$, the total magnetization $M$, the entropy $S$ and the magnetic Gr\"uneisen prameter $\Gamma_H$ of the investigated spin-electron tetrahedral chain. The first three physical quantities immediately follow from the relation~(\ref{eq:DIT}):
\begin{eqnarray}
\label{eq:G}
{\cal G}\!\!\!&=&\!\!\!-\frac{1}{\beta}\ln{\cal Z}_{I} - \frac{N}{\beta}\ln A\,,
\\
\label{eq:M}
M \!\!\!&=&\!\!\! M_I + 2M_e= -\frac{\partial {\cal G}}{\partial H_I} -2\frac{\partial {\cal G}}{\partial H_e}
\nonumber \\
        &=&\!\!\!
\frac{1}{\beta}\frac{\partial \ln{\cal Z}_{I}}{\partial H_I}+ \frac{N}{\beta}\frac{\partial \ln A}{\partial H_I}+ \frac{2}{\beta}\frac{\partial \ln{\cal Z}_{I}}{\partial H_e}+ \frac{2N}{\beta}\frac{\partial \ln A}{\partial H_e}\,,\nonumber \\
\\
\label{eq:S}
S \!\!\!&=&\!\!\!-\frac{\partial {\cal G}}{\partial T}= \beta^2\frac{\partial {\cal G}}{\partial \beta}\nonumber \\
        &=&\!\!\!
\ln{\cal Z}_{I} + N\ln A - \beta\frac{\partial \ln{\cal Z}_{I}}{\partial \beta}- N\beta\frac{\partial \ln A}{\partial \beta}\,.
\end{eqnarray}
The partial derivatives of the functions $\ln{\cal Z}_{I}$ and $\ln A$, appearing in Eqs.~(\ref{eq:M}) and~(\ref{eq:S}), satisfy the general equations:
\begin{eqnarray}
\label{eq:der_lnZ_IC}
\frac{\partial \ln{\cal Z}_{I}}{\partial x} \!\!\!&=&\!\!\!
 \frac{N}{2}\left[\frac{1}{2} + \frac{s^2-Q^2}{Q(c+Q)}+\frac{s}{Q}\right]\frac{\partial \ln(W_{-}\! + W)}{\partial x}
\nonumber \\
        & &\!\!\! +\frac{N}{2}\left[\frac{1}{2} + \frac{s^2-Q^2}{Q(c+Q)}-\frac{s}{Q}\right]\frac{\partial \ln(W_{+}\! + W)}{\partial x}
\nonumber \\
&&\!\!\! -N\left[\frac{1}{2} + \frac{s^2-Q^2}{Q(c+Q)}\right]\frac{\partial \ln(W_{0}\! + W)}{\partial x}
\nonumber \\
        & &\!\!\!+
\frac{Ns}{2Q}\frac{\partial (\beta H_{I})}{\partial x}\,,
\\
\label{eq:der_lnA}
\frac{\partial \ln A}{\partial x} \!\!\!&=&\!\!\!
\frac{1}{4}  \frac{\partial\ln(W_{-}\! + W)}{\partial x} +  \frac{1}{4} \frac{\partial\ln(W_{+}\! + W)}{\partial x} \nonumber \\
        & &\!\!\!+  \frac{1}{2} \frac{\partial\ln(W_{0}\! + W)}{\partial x}\,,
\end{eqnarray}
where $s = \sinh\left(\beta H_{e\!f\!f}/2\right)$, $c = \cosh\left(\beta H_{e\!f\!f}/2\right)$ and $Q = \sqrt{\sinh^2\left(\beta H_{e\!f\!f}/2\right) + \exp\left(-\beta J_{e\!f\!f}\right)}$. Forms of the functions $W_{\mp}$, $W_0$ and $W$ are listed in Eq.~(10) of Ref.~\cite{Gal15b}.

The so-called magnetic Gr\"uneisen parameter $\Gamma_H$, which can be calculated from the relation (see Ref.~\cite{Wol14} for a recent review):
\begin{eqnarray}
\label{eq:Gamma1}
\Gamma_H \!\!\!&=&\!\!\! -\frac{1}{C_H}\left(\frac{\partial M}{\partial T}\right)_H\!=  -\frac{1}{T} \frac{\left(\partial S/\partial H\right)_T}{\left(\partial S/\partial T\right)_H} \!= \frac{1}{T}\left(\frac{\partial T}{\partial H}\right)_S
\end{eqnarray}
($C_H$ is the specific heat at the constant magnetic field $H$), refers to a thermal response of magnetic system with respect to a variation of the external magnetic field. This physical quantity represents a magnetic analog of the classical thermal Gr\"uneisen parameter~\cite{Gru12,Zhu03,Gar05}. It is noteworthy that the magnetic Gr\"uneisen parameter~(\ref{eq:Gamma1}) diverges at a quantum phase transition driven by the external magnetic field quite similarly as the thermal Gr\"uneisen parameter does at a quantum phase transition driven by the external pressure. From this point of view, the magnetic Gr\"uneisen parameter~(\ref{eq:Gamma1}) provides a valuable tool for an experimental identification of the field-tuned quantum phase transitions~\cite{Geg10,Wei12,Ryl14}.
In addition, it is clear from Eq.~(\ref{eq:Gamma1}) that the  magnetic Gr\"uneisen parameter $\Gamma_H$ is proportional to the adiabatic cooling rate $\left(\partial T/\partial H\right)_S$ and hence, it represents a key physical quantity for an investigation of the cooling efficiency  during the adiabatic demagnetization especially in a vicinity of field-induced phase transitions. A~direct substitution of the total magnetization~(\ref{eq:M}) and the temperature derivative of the entropy~(\ref{eq:S}) into the expression~(\ref{eq:Gamma1}) yields the following form of the magnetic Gr\"uneisen parameter for the hybrid spin-electron double-tetrahedral chain:
\begin{eqnarray}
\label{eq:Gamma2}
\Gamma_H
 \!\!\!&=&\!\!\!
-\frac{\left(\frac{\partial \ln{\cal Z}_{I}}{\partial H_I}\right)+ N\left(\frac{\partial \ln A}{\partial H_I}\right)+ 2\left(\frac{\partial \ln{\cal Z}_{I}}{\partial H_e}\right)+ 2N\left(\frac{\partial \ln A}{\partial H_e}\right)}{\beta^2\left(\frac{\partial^2 \ln{\cal Z}_{IC}}{\partial \beta^2}\right)+ N\beta^2\left(\frac{\partial^2 \ln A}{\partial \beta^2}\right)}
\nonumber \\
        & &\!\!\!+
\frac{\left(\frac{\partial^2 \ln{\cal Z}_{I}}{\partial H_I\partial\beta}\right)+ N\left(\frac{\partial^2 \ln A}{\partial H_I\partial\beta}\right)+ 2\left(\frac{\partial^2 \ln{\cal Z}_{I}}{\partial H_e\partial\beta}\right)+ 2N\left(\frac{\partial^2 \ln A}{\partial H_e\partial\beta}\right)}{\beta\left(\frac{\partial^2 \ln{\cal Z}_{I}}{\partial \beta^2}\right)+ N\beta\left(\frac{\partial^2 \ln A}{\partial \beta^2}\right)}\,.
\end{eqnarray}
The second partial derivatives of the functions $\ln{\cal Z}_{I}$, $\ln A$ that emerge in Eq.~(\ref{eq:Gamma2}) can be obtained by differentiating Eqs.~(\ref{eq:der_lnZ_IC}) and~(\ref{eq:der_lnA}) with respect to the relevant variable. It should be noted, however, that the resulting expressions for these derivatives are too cumbersume to write them here explicitly.

\section{Results and discussion}
\label{sec:3}
In this section, we present the most interesting results obtained for the spin-electron double-tetrahedral chain by considering the particular case with the ferromagnetic Ising interaction $J<0$ between the localized Ising spins and mobile electrons. To reduce the number of free interaction parameters, we will also assume equal 'effective' magnetic fields $H = H_{ I}= H_{e}$ acting on magnetic particles.

\subsection{Ground state}
\label{subsec:GS}
\begin{figure}[ht!]
\begin{center}
\vspace{-0.25cm}
\includegraphics[angle = 0, width = 0.85\columnwidth]{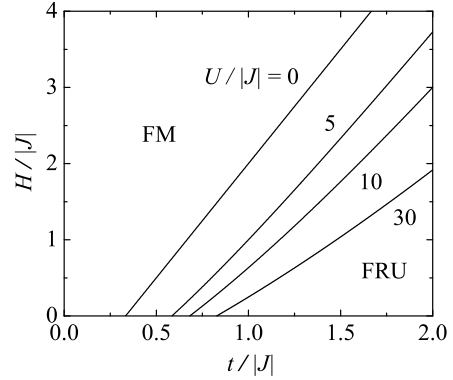}
\vspace{-0.25cm}
\caption{\small The ground-state phase diagram of the spin-electron double-tetrahedral chain with the ferromagnetic coupling $J<0$ constructed in the $t/|J|-H/|J|$ plane for the  Coulomb terms $U/|J|=0, 5, 10, 30$.}
\label{fig2}
\end{center}
\vspace{-0.25cm}
\end{figure}

First, let us comment on possible phases that may appear in the ground state of the investigated spin-electron model. A typical ground-state phase diagram, constructed in the $t/|J| - H/|J|$ plane by assuming various values of the Coulomb term $U/|J|$, is displayed in Fig.~\ref{fig2}. As one can see from this figure, the ground state of the system contains two possible phases, namely, the ferromagnetic (FM) phase and the frustrated (FRU) phase. The boundary, which represents the first-order transition between the relevant phases, is given by the condition:
\begin{eqnarray}
\label{eq:H_c}
\frac{H_c}{|J|}
 \!\!\!&=&\!\!\!  \sqrt{\left(\frac{U}{2|J|}+\frac{t}{|J|}\right)^2+8\left(\frac{t}{|J|}\right)
^2} - \frac{U}{2|J|} -1\,.
\label{eq:FRUFM}
\end{eqnarray}
It is worthwhile to remark that the FM and FRU phases also appear in the ground state of the antiferromagnetic counterpart of the model (see our preceding paper~\cite{Gal15b}). Thus, we present here just their eigenvectors and brief definitions for the sake of easy reference:
\begin{itemize}
\item The ferromagnetic (FM) phase:
\begin{eqnarray}
\label{eq:FM}
\hspace{-1.0cm}
      |{\rm FM}\rangle = \prod_{k=1}^N
        |\!\uparrow\rangle_{\sigma_k}\!\otimes\!
        \begin{cases}
        \lefteqn{\frac{1}{\sqrt{3}}\,
(c_{k1,\uparrow}^{\dag}c_{k2,\uparrow}^{\dag}\! +\omega c_{k2,\uparrow}^{\dag}c_{k3,\uparrow}^{\dag}\! +\omega^{2} c_{k3,\uparrow}^{\dag}c_{k1,\uparrow}^{\dag})|0\rangle}
                \\[3mm]
       \lefteqn{\frac{1}{\sqrt{3}}\,
(c_{k1,\uparrow}^{\dag}c_{k2,\uparrow}^{\dag}\! +\omega^{2} c_{k2,\uparrow}^{\dag}c_{k3,\uparrow}^{\dag}\! +\omega c_{k3,\uparrow}^{\dag}c_{k1,\uparrow}^{\dag})|0\rangle,}
         \end{cases}\nonumber\\
    \end{eqnarray}
where  $\omega = {\rm e}^{2\pi {\rm i}/3}$ (${\rm i}^2=-1$). In this phase, the mobile electrons at each triangular cluster underlie a quantum superposition of three ferromagnetic states $c_{k1,\uparrow}^{\dag}c_{k2,\uparrow}^{\dag}|0\rangle$, $c_{k2,\uparrow}^{\dag}c_{k3,\uparrow}^{\dag}|0\rangle$, $c_{k3,\uparrow}^{\dag}c_{k1,\uparrow}^{\dag}|0\rangle$ and the Ising spins localized at nodal lattice sites occupy the spin state $\sigma^z=1/2$.

\item The frustrated (FRU) phase:
\begin{eqnarray}
\label{eq:FRU}
\hspace{-0.75cm}
|{\rm FRU}\rangle=\begin{cases}
                    \lefteqn{{}\prod\limits_{k=1}^N |\!\uparrow\!(\downarrow)\rangle_{\sigma_k}\!\otimes\!
                          \frac{1}{\sqrt{6}}\,\big[\sin\varphi\,(c_{k1,\uparrow}^{\dag}c_{k2,\downarrow}^{\dag}\!+ c_{k2,\uparrow}^{\dag}c_{k3,\downarrow}^{\dag} {}}
                          \nonumber\\
                          \lefteqn{\hspace{0.75cm}{}
													 +c_{k3,\uparrow}^{\dag}c_{k1,\downarrow}^{\dag} \!- c_{k1,\downarrow}^{\dag}c_{k2,\uparrow}^{\dag} \!- c_{k2,\downarrow}^{\dag}c_{k3,\uparrow}^{\dag} \!- c_{k3,\downarrow}^{\dag}c_{k1,\uparrow}^{\dag})}
													\nonumber\\
                          \lefteqn{\hspace{0.75cm}{}
							             +  \sqrt{2}\cos\varphi\sum\limits_{j=1}^{3}c_{kj,\uparrow}^{\dag}c_{kj,\downarrow}^{\dag}\big]|0\rangle \quad (H=0)}\\
                          \lefteqn{{}\prod\limits_{k=1}^N|\!\uparrow\rangle_{\sigma_k}\!\otimes\!
                           \frac{1}{\sqrt{6}}\,\big[\sin\varphi\,(c_{k1,\uparrow}^{\dag}c_{k2,\downarrow}^{\dag}\!+ c_{k2,\uparrow}^{\dag}c_{k3,\downarrow}^{\dag} }
													\end{cases}
                          \nonumber\\
                          \lefteqn{\hspace{0.15cm}{}
													+ c_{k3,\uparrow}^{\dag}c_{k1,\downarrow}^{\dag} \!- c_{k1,\downarrow}^{\dag}c_{k2,\uparrow}^{\dag} \!-c_{k2,\downarrow}^{\dag}c_{k3,\uparrow}^{\dag} \!
                                        - c_{k3,\downarrow}^{\dag}c_{k1,\uparrow}^{\dag})}
                        \nonumber\\
                                 \lefteqn{\hspace{0.25cm}+  \sqrt{2}\cos\varphi\sum\limits_{j=1}^{3}c_{kj,\uparrow}^{\dag}c_{kj,\downarrow}^{\dag}\big]|0\rangle \quad  (H>0),}
\end{eqnarray}
where $\tan\varphi = \frac{\sqrt{2}}{8t}\left(U+2t+\sqrt{(U+2t)^2+32t^2}\,\right)$. In this phase, the mobile electrons from each triangular cluster show the quantum entanglement of six intrinsic antiferromagnetic states $c_{k1,\uparrow}^{\dag}c_{k2,\downarrow}^{\dag}|0\rangle$, $c_{k2,\uparrow}^{\dag}c_{k3,\downarrow}^{\dag}|0\rangle,  c_{k3,\uparrow}^{\dag}c_{k1,\downarrow}^{\dag}|0\rangle,  c_{k1,\downarrow}^{\dag}c_{k2,\uparrow}^{\dag}|0\rangle, c_{k2,\downarrow}^{\dag}c_{k3,\uparrow}^{\dag}|0\rangle, c_{k3,\downarrow}^{\dag}c_{k1,\uparrow}^{\dag}|0\rangle$ and three non-magnetic ionic states $c_{kj,\uparrow}^{\dag}c_{kj,\downarrow}^{\dag}|0\rangle$ ($j=1,2,3$), while the arrangement of the localized Ising spins depends on a presence of the external magnetic field: if $H=0$, the Ising spins are completely free to flip in arbitrary direction, while they are fully polarized into the field direction, if $H\neq 0$.
\end{itemize}
In both Eqs.~(\ref{eq:FM}) and~(\ref{eq:FRU}), the products run over all primitive unit  cells, the state vector $|\!\uparrow\rangle_{\sigma_k}$ ($|\!\downarrow\rangle_{\sigma_k}$) determines the up (down) state of the Ising spin localized at $k$th lattice site and $|0\rangle$ labels the vacuum state. As has been evidenced in Ref.~\cite{Gal15b}, the above ground states are macroscopically degenerate, which consequently leads to a residual entropy $S/3N = \ln 2^{1/3}\approx 0.231$ in both the phases. The FM phase is macroscopically degenerate due to chiral degrees of freedom of the mobile electrons, while the FRU phase exhibits a macroscopic degeneracy owing to a kinetically-driven spin frustration of the localized Ising spins caused by the antiferromagnetic alignment of the mobile electrons. Arbitrary but non-zero magnetic field tends to align the Ising spins into the field direction and thus, it cancels the macroscopic degeneracy of the FRU phase (and also the associated residual entropy  $S/3N = \ln 2^{1/3}$). By contrast, the FM phase remains macroscopically degenererate in the whole parameter region.

\subsection{Adiabatic (de)magnetization process}
\label{subsec:SGamma}

\begin{figure}[h!]
\begin{center}
\vspace{-0.25cm}
\includegraphics[angle = 0, width = 0.85\columnwidth]{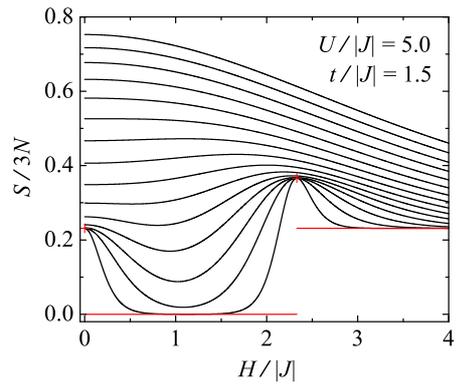}
\vspace{-0.25cm}
\caption{\small The isothermal dependencies of the entropy $S/3N$ on the magnetic field $H/|J|$ for the fixed Coulomb term $U/|J|=5.0$, the hopping parameter $t/|J|=1.5$ and the various temperatures $T/|J|=0, 0.1, 0.2,\ldots, 1.5$ (from bottom to top).}
\label{fig3}
\end{center}
\vspace{-0.25cm}
\end{figure}

Now, let us turn to the discussion of the magnetocaloric properties of the investigated model. In Fig.~\ref{fig3} we depict isothermal changes of the entropy per one magnetic particle $S/3N$ (recall that the system is composed of $N$ Ising spins and $2N$ mobile electrons) under the field variation assuming the fixed Coulomb term $U/|J|=5.0$, the hopping parameter $t/|J|=1.5$ and various temperatures. As one can see from this figure, the entropy isotherms monotonously decrease from its maximum at $H/|J|=0$ upon the increasing magnetic field down to temperature $T/|J|=1.0$. Below $T/|J|=1.0$, the entropy exhibits non-monotonous dependencies as a function of the external magnetic field with a pronounced peak at the critical field~(\ref{eq:H_c}), at which the system undergoes a phase transition between the FRU and FM phases. Finally, the continuous entropy isotherms split
into the isolated points with the coordinates $[H/|J|,S/3N]=[0, \ln 2^{1/3}]$, $[H/|J|,S/3N]=[H_c/|J|, \ln 3^{1/3}]$ and the lines $S/3N=0$ ($H<H_c$), $S/3N=\ln 2^{1/3}$ ($H>H_c$) when the temperature reaches the zero value.

\begin{figure}[h!]
\begin{center}
\vspace{-0.25cm}
\includegraphics[angle = 0, width = 0.85\columnwidth]{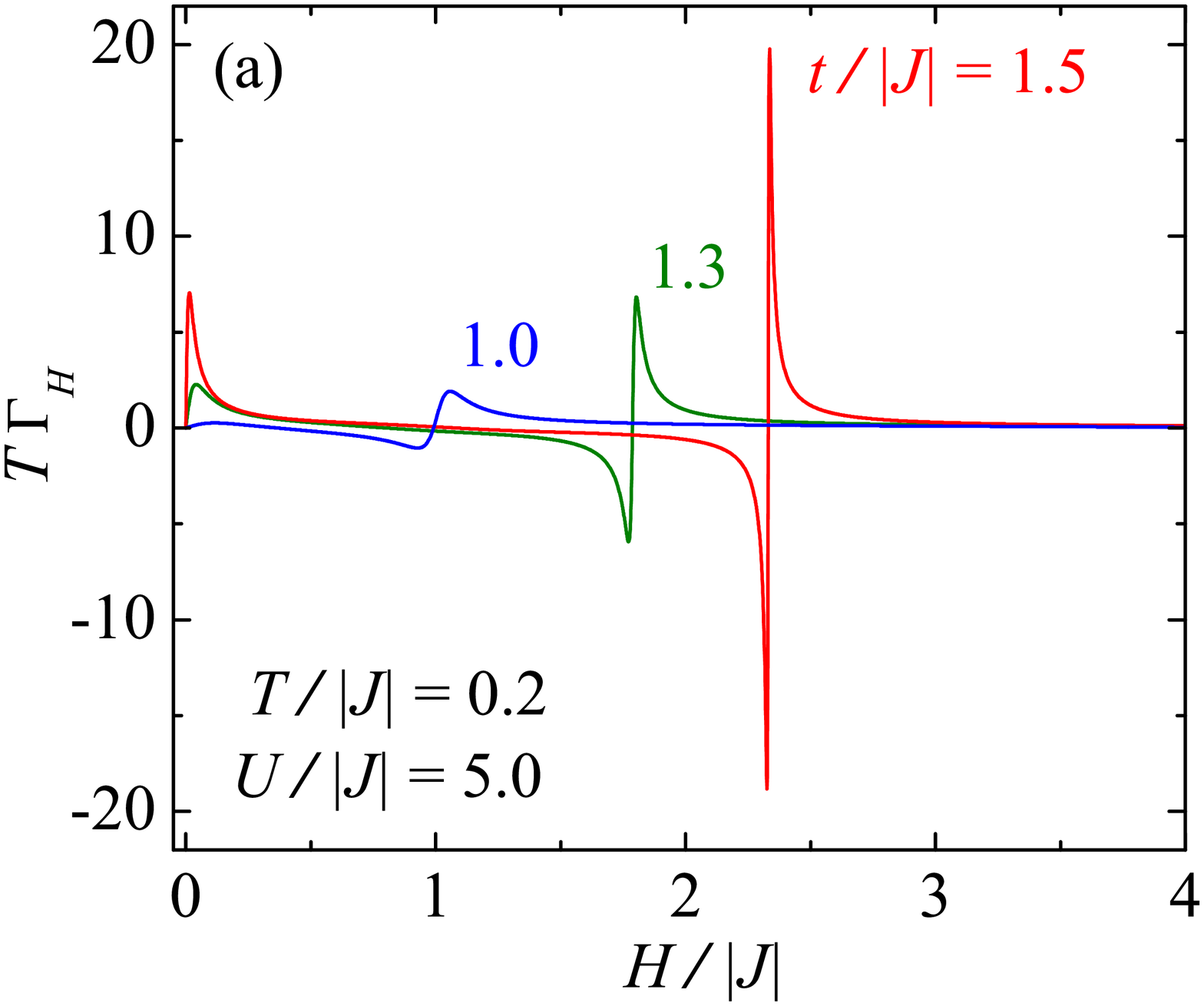}\\[-5mm]
\includegraphics[angle = 0, width = 0.85\columnwidth]{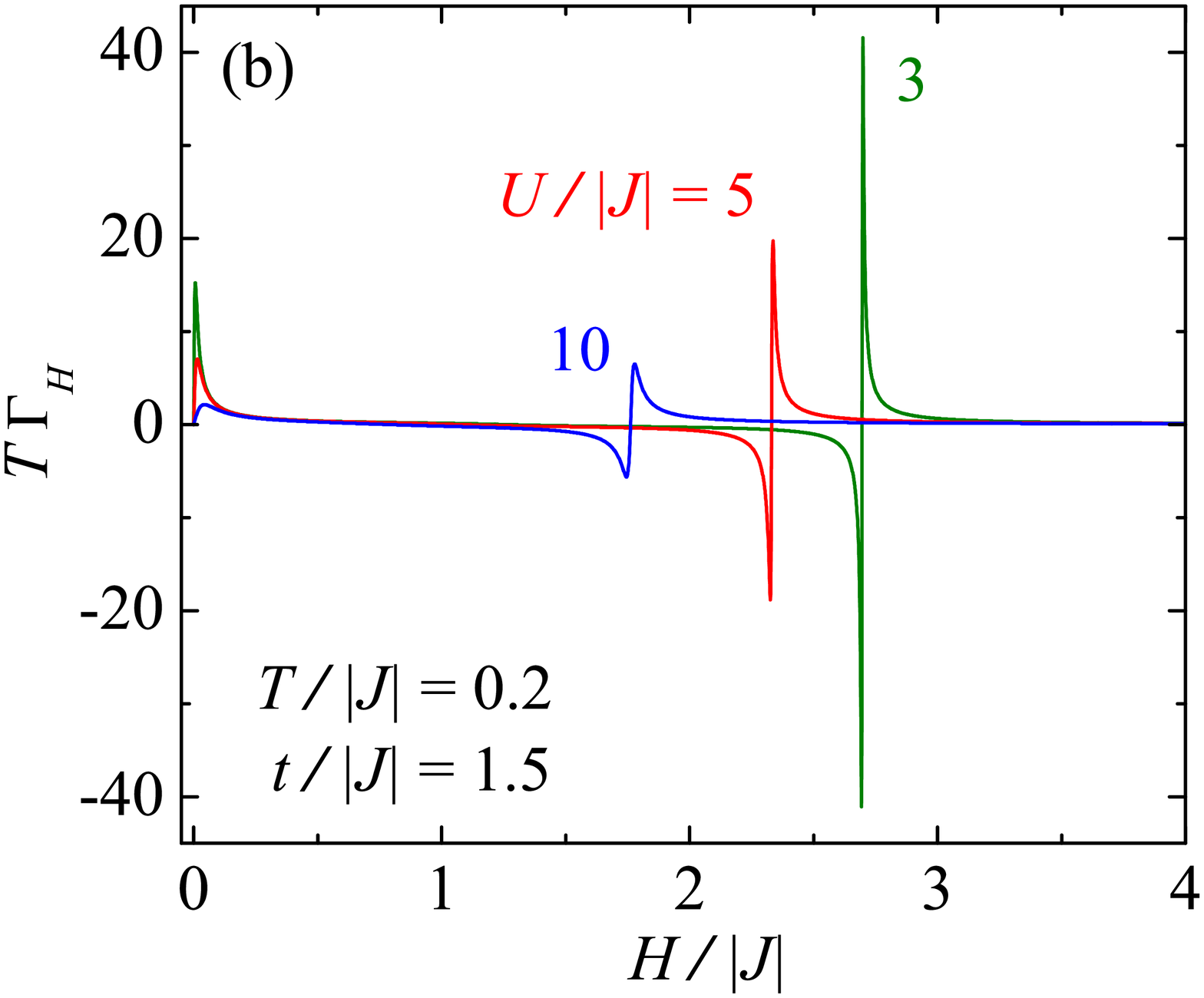}
\vspace{-0.25cm}
\caption{\small The magnetic Gr\"uneisen parameter multiplied by the temperature $T\Gamma_H$ versus the magnetic field for the fixed temperature $T/|J|=0.2$ and the set of parameters: (a) $U/|J|=5.0$, $t/|J|=1.0, 1.3, 1.5$, (b) $t/|J|=1.5$, $U/|J|=3, 5, 10$.}
\label{fig4}
\end{center}
\vspace{-0.25cm}
\end{figure}

In general, the residual entropy found in the frustrated magnetic systems and at critical points corresponding to the phase transitions between different magnetic structures gives a rise to an enhanced MCE accompanied by a relatively fast cooling of systems during the adiabatic (de)magnetization. To investigate the efficiency of the adiabatic cooling of the considered spin-electron system in these regions, we plot in Fig.~\ref{fig4} the magnetic Gr\"uneisen parameter multiplied by the temperature $T\Gamma_H$ versus the external magnetic field for the relatively low temperature $T/|J| = 0.2$ and a few values of the parameters $t/|J|$ and $U/|J|$. Evidently, the low-temperature $T\Gamma_H$ curves depicted in this figure exhibit two local maxima at very low (but non-zero) magnetic field and slightly above the critical field~(\ref{eq:H_c}) of the zero-temperature phase transition FRU--FM in addition to the one local minimum slightly below~(\ref{eq:H_c}). The sign change of the $T\Gamma_H$ product observed close to the critical field~(\ref{eq:H_c}) clearly indicates a rapid accumulation of the entropy due to a mutual competition between the neighbouring ground-state configurations (see e.g.~the curves plotted for $t/|J|=1.5$ in Fig.~\ref{fig4}a and for $U/|J|=5$ in Fig.~\ref{fig4}b and compare them with the corresponding isothermal dependence of $S/3N$ in Fig.~\ref{fig3}). Furthermore, the peaks appearing around the field-induced phase transition
FRU--FM are much higher than those observed at relatively small magnetic fields.
For example, for $t/|J|=1.5$ and $U/|J|=3, 5, 10$ we have $T\Gamma_H\approx 41.6, 19.8,6.5$ at $H/|J|\approx 2.7, 2.3,1.8$ and $T\Gamma_H\approx 15.3, 7.1, 2.1$ at $H/|J|\approx 0.007, 0.014, 0.043$ (see Fig.~\ref{fig4}b). It is thus clear that the cooling capability of the system during the adiabatic (de)magnetization is substantially higher (approximately three times) just above the critical field, where strong thermal excitations of the mobile electrons are present at relatively low  temperatures due to breaking up the quantum superpositions of their up-down states, than that one at relatively small magnetic fields, where the Zeeman's splitting of energy levels of the frustrated Ising spins takes place.
Except to this behaviour, Fig.~\ref{fig4} also illustrates the effect of the hopping parameter and the on-site Coulomb repulsion on the enhancement of the MCE in the investigated model. It is quite evident from this figure that the adiabatic cooling rate of the system generally increases in the vicinity of the zero field and nearby the field-induced phase transition FRU--FM with increasing the kinetic term $t/|J|$ (see Fig.~\ref{fig4}a), while it decreases in these regions with increasing the Coulomb parameter $U/|J|$ (see Fig.~\ref{fig4}b).

\begin{figure}[h!]
\begin{center}
\vspace{-0.25cm}
\includegraphics[angle = 0, width = 0.85\columnwidth]{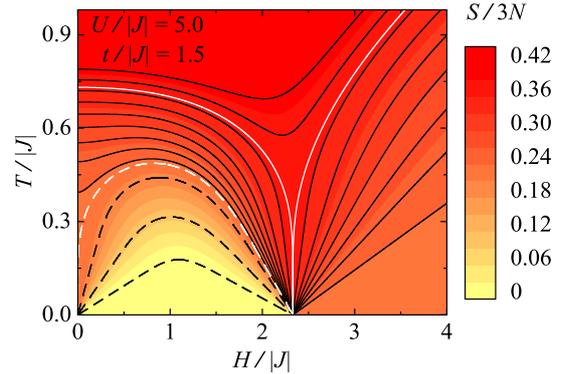}
\vspace{-0.25cm}
\caption{\small  A density plot of the entropy as a function of the magnetic field and temperature by assuming the fixed Coulomb term $U/|J|$ = 5.0 and the hopping term $t/|J| = 1.5$. The displayed curves correspond to isentropy lines, namely, $S/3N = 0.01, 0.1, 0.2$ (black broken curve), $S/3N = \ln 2^{1/3}$ (white broken curve), $S/3N = 0.24, 0.26, \ldots, 0.4$ (black solid curves) and $S/3N = \ln 3^{1/3}$ (white solid curve).}
\label{fig5}
\end{center}
\vspace{-0.25cm}
\end{figure}

To discuss the MCE in the considered spin-electron double-tetrahedral chain, we may alternatively investigate an adiabatic change of temperature of the model under the magnetic field variation. For this purpose, we present a density plot of the entropy as a function of the magnetic field and temperature for the fixed Coulomb term $U/|J|=5.0$ and the hopping parameter $t/|J|=1.5$ in Fig.~\ref{fig5}. Isentropic changes of temperature upon varying the magnetic field can be identified in this figure as contours of the constant entropy displayed by solid and broken lines. In accordance with the previous discussion, the enhanced MCE during the adiabatic (de)magnetization can be found just above the zero field as well as nearby the critical field corresponding to the field-induced phase transition between the FRU and FM phases. The most abrupt drop in temperature up to the zero value is achieved under the adiabatic condition in these regions if the entropy of the system
is set sufficiently close to the values $S/3N = \ln 2^{1/3}\approx 0.231$ and $S/3N = \ln 3^{1/3}\approx 0.366$, respectively.

\section{Conclusions}
\label{sec:4}

The present Letter deals with the ground-state and magnetocaloric properties of the double-tetrahedral chain, where the nodal lattice sites occupied by the localized Ising spins regularly alternate with three equivalent lattice sites available for two mobile electrons. Based on the exact solution of the model presented in Ref.~\cite{Gal15b}, we have analytically derived exact results for the basic thermodynamic quantities, such as the Gibbs free energy, the total magnetization, the entropy and the magnetic Gr\"uneisen parameter of the system. By considering the ferromagnetic exchange interaction ($J<0$) between the mobile electrons and their nearest Ising neighbours, we have found two macroscopically degenerate ground states (FRU and FM), which can also be found in the antiferromagnetic counterpart of the model~\cite{Gal15b}. We have shown that the macroscopic degeneracy of the FRU phase arising due to the kinetically-driven frustration of the localized Ising spins as well as the macroscopic degeneracy of the system at the field-induced phase transition between the FRU and FM phases perfectly manifest themselves in the enhanced MCE during the adiabatic (de)magnetization. It has been evidenced that the cooling capability of the system during the adiabatic (de)magnetization is approximately three times higher nearby the phase transition FRU--FM, where strong thermal excitations of the mobile electrons are present at relatively low temperatures due to breaking up the quantum superpositions of their up-down states, than that one at relatively small magnetic fields, where the Zeeman's splitting of energy levels of the frustrated Ising spins takes place.

\end{document}